\documentclass[12pt]{article}

\usepackage[utf8]{inputenc}
\usepackage[T1]{fontenc}
\usepackage[english]{babel}
\usepackage{amssymb}
\usepackage{mathtools}
\usepackage{dsfont}
\usepackage{bm}
\usepackage{braket}
\usepackage{simplewick}
\usepackage{slashed}
\usepackage{array}
\usepackage{dcolumn}
\usepackage{longtable}
\usepackage{rotating}
\usepackage{graphicx}
\usepackage{graphicx}
\usepackage{float}
\usepackage{geometry}
\usepackage{xcolor}
\DeclareMathOperator{\arctanh}{arctanh}
\usepackage[colorlinks]{hyperref}

\definecolor{color_01}{rgb}{0,0.1,0.75}
\definecolor{color_02}{rgb}{0,0.35,0.65}
\definecolor{color_03}{rgb}{0,0.35,0.65}
\hypersetup{linktoc=page,linkcolor=color_01,citecolor=color_02,urlcolor=color_03}

\usepackage{bookmark}
\usepackage{float}
\usepackage{amsmath}
\date{}
\usepackage{epstopdf}
\begin{document}

\title{Geodesic Structure of the Accelerated \\ Stephani Universe}
\maketitle
\begin{center}
Irina Bormotova $^{1, 2, *}$, Elena Kopteva $^{1}$ and Zden\v{e}k Stuchl\'{\i}k $^{1}$ \\   
    $^{1}$ \quad Research Centre for Theoretical Physics and Astrophysics, Institute of Physics, Silesian University in Opava, CZ-746~01 Opava, Czech Republic;\\
	$^{2}$ \quad Bogoliubov Laboratory of Theoretical Physics, Joint Institute for Nuclear Research, 141980~Dubna, Russia\\
	$^*${e-mail: q\_leex@mail.ru}\\
\end{center}

\begin{abstract}
For the spherically symmetric Stephani cosmological model with an accelerated expansion, we investigate the main scenarios of the test particle and photon motion. We show that a comoving observer sees an appropriate picture. In the case of purely radial motion, the radial velocity decreases slightly with time due to the universe expansion. Both particles and photons spiral out of the center when the radial coordinate is constant. In the case of the motion with arbitrary initial velocity, the observable radial distance to the test particle can increase under negative observable radial velocity.
\end{abstract}

\section{Introduction}
The $\Lambda$CDM model, based on the Friedmann solution, is commonly accepted in the  description of the observed universe's accelerated expansion. However, some crucial problems of this model, like the problem of ``dark energy'' and the coincidence problem \cite{Weinberg89}, are still unsolved. Therefore, alternative approaches are of a certain interest in cosmology. One of the possibilities here is to consider inhomogeneous cosmological models. The Stephani solution \cite{Stephani67} is among them. It allows building of the model of the universe with accelerated expansion within general relativity with no modifications or suggestions of the exotic types of matter \cite{Borm20,Dabrowski98,  Stelmach04, Stelmach08, Balcerzak15, Ong18}. This is a non-static solution for expanding perfect fluid with zero shear and rotation, which contains the known Friedmann solution as a particular case. The Stephani solution was discussed extensively in the literature (see e.g., \cite{Ong18, Krasinski83, Sussman87, Sussman88a, Sussman88b, Sussman00, Dabrowski93, Korkina16} and references therein). Initially, it has no symmetries, but the spatial sections of the Stephani space--time in the case of spherical symmetry have the same geometry as corresponding subspaces of the Friedmann solution. Therefore, these models have an intuitively clear interpretation. The spatial curvature in the Stephani solution depends on time that allows the attainment of the accelerated expansion of the universe. 

Recently, a particular case of the Stephani solution was investigated as a possible model of the accelerated universe \cite{Kopteva19}. It was shown that the theoretical prediction for the redshift--magnitude relation in the model is in good accordance with type Ia Supernovae~data.

In this paper, we investigate the geodesic structure of the model obtained in \cite{Kopteva19} to verify if the singularities of the model can affect the light and test particle motion within the observable area.

The paper is organized as follows. In Section~\ref{Sec2} we introduce the special case of the Stephani solution for the universe with accelerated expansion. In Section~\ref{Sec3} the geodesic equations for the considered model are investigated. The null geodesics are studied in Section~\ref{Sec4}. The conclusions are presented in Section~\ref{Con}.

\section{The Accelerated Stephani Universe}\label{Sec2}
Let us consider the inhomogeneous spherically symmetric universe filled with a perfect fluid with uniform energy density $\varepsilon(t)$ and non-uniform pressure $p(t,\chi)$. 

The perfect fluid is described by the four-velocity vector field $u^\alpha$ with special properties that impose restrictions on the model:
\begin{eqnarray}\label{thetadef}
&\Theta &=u^{\alpha}_{;\alpha}, \\ 
&\omega_{\alpha\beta}&=u_{[\alpha;\beta]}+\dot{u}_{[\alpha}u_{\beta]}, \\
&\sigma_{\alpha\beta}&=u_{(\alpha;\beta)}+\dot{u}_{(\alpha}u_{\beta)}-\Theta h_{\alpha\beta}/3,\\
&h_{\alpha\beta}&=g_{\alpha\beta}+u_{\alpha}u_{\beta},
\end{eqnarray}
where $\dot{u}_{\alpha}$ is the acceleration, $\Theta$ is the volume expansion, $\omega_{\alpha\beta}$ is the rotation, $\sigma_{\alpha\beta}$ is the shear, $h_{\alpha\beta}$ is the projection tensor, Greek indexes run from zero to three, square/round brackets mean antisymmetrization/symmetrization by corresponded indexes (see e.g., \cite{Stephani03}). Here and further dot denotes the partial derivative with respect to time, and we take a geometric unit system $c \equiv 1$, $ 8\pi G \equiv 1$. Sometimes in the paper we shall skip the variable of the function in the formulae if it does not lead to confusion.

The model of the universe under consideration is described by the Stephani solution written in comoving coordinates \cite{Kopteva19}:
\begin{equation}\label{ourm2}
\mathrm{d}s^2=\frac{\dot{r}^2(t,\chi) a^2(t)}{r^2(t,\chi) \dot{a}^2(t)}\mathrm{d}t^2-r^2(t,\chi)\left(\mathrm{d}\chi^2+\chi^2\mathrm{d}\Omega^2\right),
\end{equation}
where
\begin{equation}\label{rfin}
r(t,\chi)=\frac{a(t)}{1+\zeta(t) a^2(t) \left(\frac{\chi}{2}\right)^2},
\end{equation}
\begin{equation}\label{zeta}
\zeta(t)=-|\beta|\frac{a_0^k(t)}{a^{k+2}(t)},
\end{equation}
$\mathrm{d}\Omega^2$ is the usual metric on the unit 2-sphere, $\zeta(t)$ is the spatial curvature, $a(t)$ is an arbitrary function related to the scale factor of the Friedmann solution, $k=\mathrm{const}~<-1$, $\beta=~\mathrm{const}<0$, which means that the spatial curvature is negative everywhere in the~universe.  
  
Here, the four-velocity of fluid particles has the following components:
\begin{eqnarray}
&u^0&=\frac{1}{\sqrt{g_{00}}},\\
&u^i&=0, \quad i=1,2,3,
\end{eqnarray} 
and the shear $\sigma_{\alpha\beta}$ and the rotation $\omega_{\alpha\beta}$ are zero.   

In view of (\ref{rfin}), $g_{00}$ takes the following explicit form
\begin{equation}\label{g00}
g_{00}=\frac{\left(1-|\beta|(k+1)\left(\frac{a_0}{a(t)}\right)^k \frac{\chi^2}{4}\right)^2}{\left(1-|\beta|\left(\frac{a_0}{a(t)}\right)^k \frac{\chi^2}{4}\right)^2}.
\end{equation}

The main equation that governs the evolution of the model reads
\begin{equation}\label{maineq}
\frac{\dot{a}^2(t)}{a^2(t)}+\zeta(t)=\varepsilon(t). 
\end{equation}

The energy density in the model has the form of the Friedmann dust: 
\begin{equation}\label{enden}
\varepsilon(t) =\frac{a_0}{a^3(t)},
\end{equation}
$a_0=\mathrm{const}=a(t_0)$, where $t_0$ corresponds to the current moment of time (our time).

The pressure is given by
\begin{equation}
p(t,\chi)=\frac{a_0}{a^3(t)} \frac{\left(\frac{\chi}{2}\right)^2|\beta|k}{\left(\frac{a(t)}{a_0}\right)^k-\left(\frac{\chi}{2}\right)^2|\beta|(k+1)}.
\end{equation}

It was shown \cite{Dabrowski98, Krasinski83, Sussman88b} that the Stephani models contain some special singularities that should be taken into account if one intends to build a cosmological model.
In this study, the model contains three true singularities.  
\begin{enumerate}
\item The initial singularity (the Big Bang): $a(t)=0$ $\Rightarrow$ $r=0$, $\varepsilon \to \infty$, $p \to \infty$.
\item The singularity arising from $g_{11}$:
\begin{equation}\label{sing}
\chi=\frac{2\left( \frac{a(t)}{a_0} \right)^{\frac{k}{2}}}{\sqrt{|\beta|}}.
\end{equation}
\item The singularity arising from $g_{00}$ affecting also the expression for pressure $p$:
\begin{equation}\label{sing2}
\chi=\frac{2\left( \frac{a(t)}{a_0} \right)^{\frac{k}{2}}}{\sqrt{(1+k)|\beta|}}.
\end{equation}
In the case of $k=-1$ the singularity points (\ref{sing2}) belong to the spatial infinity independently on the value of the time coordinate. This particular case is called the Stephani--Dabrowski model in the literature \cite{Stelmach04,  Stelmach08, Dabrowski93}. If one chooses here $k<-1$ (as we do in this paper), the singular behavior of the pressure disappears.
\end{enumerate}

The spatial sections of the space--time with metric (\ref{ourm2}) are represented by the three-dimensional hypersurface $t=t_0=\mathrm{const}$ that yields $a=a_0=\mathrm{const}$ with the intrinsic~metric 
\begin{equation}\label{dl1}
\mathrm{d}l^2=\frac{a_0^2}{\left(1-\frac{\chi^2}{4}|\beta|\right)^2} \left(\mathrm{d}\chi^2+\chi^2 \mathrm{d}\varphi^2\right),
\end{equation}
where the equatorial plane $\theta=\pi/2$ is chosen for the simplicity of geometric visualization.
By use of a new coordinate $\rho=\frac{\sqrt{|\beta|}}{2}\chi$, the metric (\ref{dl1}) can be rewritten in more familiar way
\begin{equation}\label{dl2}
\mathrm{d}l^2=\frac{a_0^2}{|\beta|}\frac{4}{\left(1-\rho^2\right)^2} \left(\mathrm{d}\rho^2+\rho^2 \mathrm{d}\varphi^2\right).
\end{equation}

This is a metric of the pseudo-sphere in terms of the stereographic projection coordinates (see e.g., \cite{Dubrovin}) accurate within the similarity transformation with constant factor $a_0^2/|\beta|$. This stereographic projection maps the upper half of the pseudo-sphere represented by the hyperboloid of revolution onto the open disk $\rho^2=x^2 + y^2 < 1$ on the plane $z=0$. Such a disk equipped with normalized metric (\ref{dl2}) (so that $a_0^2/|\beta|=1$) refers to the Poincare model of Lobachevsky geometry. It is seen that the spatial sections of the interval (\ref{ourm2}) are the Lobachevsky spaces.

The constants of the model adapted to the observational data \cite{WMAP} have the following values \cite{Kopteva19}:
\begin{eqnarray}
&a_0&=1.58\times 10^{26}~\mathrm{m},\\ \label{a0}
&\beta & = -0.111113,\\
&\chi_0 &=2.59906.
\end{eqnarray}
$\chi_0$ is the value of the coordinate corresponding to the current size of the universe: 
\begin{equation}\label{r0}
r_0=\int_0^{\chi_0}\frac{a(t_0)}{1+\zeta(t_0) a^2(t_0)\left( \frac{\chi}{2}\right)^2 }\mathrm{d}\chi=\frac{2a_0}{\sqrt{|\beta|}} \arctanh (\frac{\chi_0}{2}\sqrt{|\beta|})\approx4.4\times 10^{26}~\mathrm{m}.
\end{equation}

\section{Geodesic Equations}\label{Sec3}
We now derive the equations of motion of a test particle in the model from the point of view of a comoving observer, i.e., a local observer at rest relative to the averaged motion of the neighbor matter. As far as we have spherical symmetry, it is enough to consider the motion in the equatorial plane, so that $\mathrm{d}\theta/\mathrm{d}s=0$.

For the interval (\ref{ourm2})  
\begin{equation*}
\mathrm{d}s^2=g_{00}(t,\chi)\mathrm{d}t^2-r^2(t,\chi)\left(\mathrm{d}\chi^2 -\chi^2\mathrm{d}\sigma^2\right),
\end{equation*}
with
\begin{equation}\label{g00r}
g_{00}(t,\chi)=\frac{\left( 1-T^{-k}(t)\left(\frac{\chi}{2}\right)^2(1+k)|\beta| \right)^2}{\left( 1-T^{-k}(t)\left(\frac{\chi}{2}\right)^2|\beta| \right)^2},
\end{equation}
\begin{equation}\label{rr}
r(t,\chi)=\frac{a_0 T(t)}{\left( 1-T^{-k}(t)\left(\frac{\chi}{2}\right)^2|\beta| \right)},
\end{equation}
where $T(t)\equiv \frac{a(t)}{a_0}$, the geodesic equations found in the standard way read
\begin{equation}\label{geq_u0}
\frac{\mathrm{d}u^0}{\mathrm{d}s}=-\frac{1}{2g_{00}}\left((u^0)^2 \dot{g}_{00}+2\left(r\left((u^1)^2+\chi^2 (u^3)^2\right)\dot{r} + u^1u^0g'_{00}\right)\right),
\end{equation}
\begin{equation}\label{geq_u1}
\frac{\mathrm{d}u^1}{\mathrm{d}s}=-\frac{1}{2r^2}\left( -2\chi r^2 (u^3)^2 + (u^0)^2 g'_{00} + 2r\left( 2u^1 u^0 \dot{r} +\left((u^1)^2-\chi^2(u^3)^2 \right)r'  \right) \right),
\end{equation}
\begin{equation}\label{geq_u3}
\frac{\mathrm{d}u^3}{\mathrm{d}s}=-\frac{1}{\chi r}\left(2u^3\left(u^1 r+\chi\left(u^0\dot{r}+u^1 r'\right)\right)\right).
\end{equation}

Here the usual notations are used
\begin{equation}
u^0=\frac{\mathrm{d}t}{\mathrm{d}s}, \qquad u^1=\frac{\mathrm{d}\chi}{\mathrm{d}s}, \qquad u^3=\frac{\mathrm{d}\varphi}{\mathrm{d}s}.
\end{equation}

The normalization condition corresponding to the test particles ($u^{\mu}u_{\mu}=1$) gives one more useful relation
\begin{equation}\label{geq_int}
\frac{r^2}{g_{00}}\left(\frac{u^1}{u^0}\right)^2 + \frac{r^2\chi^2}{g_{00}}\left(\frac{u^3}{u^0}\right)^2=1-\frac{1}{g_{00}(u^0)^2}.
\end{equation}

The first term in (\ref{geq_int}) is the squared observable radial velocity of the particle $v_{r}=\frac{r}{\sqrt{g_{00}}}\frac{\mathrm{d}\chi}{\mathrm{d}t}$ and the second term is the squared observable orbital velocity $v_{\varphi}=\frac{r\chi}{\sqrt{g_{00}}}\frac{\mathrm{d}\varphi}{\mathrm{d}t}$ which would be measured by the set of local observers located along the particle trajectory and would be at rest relative to the surrounding medium.

\begin{figure}[t]
\centering
	\begin{minipage}{0.48\linewidth}\centering
 		\includegraphics[width=0.9\linewidth]{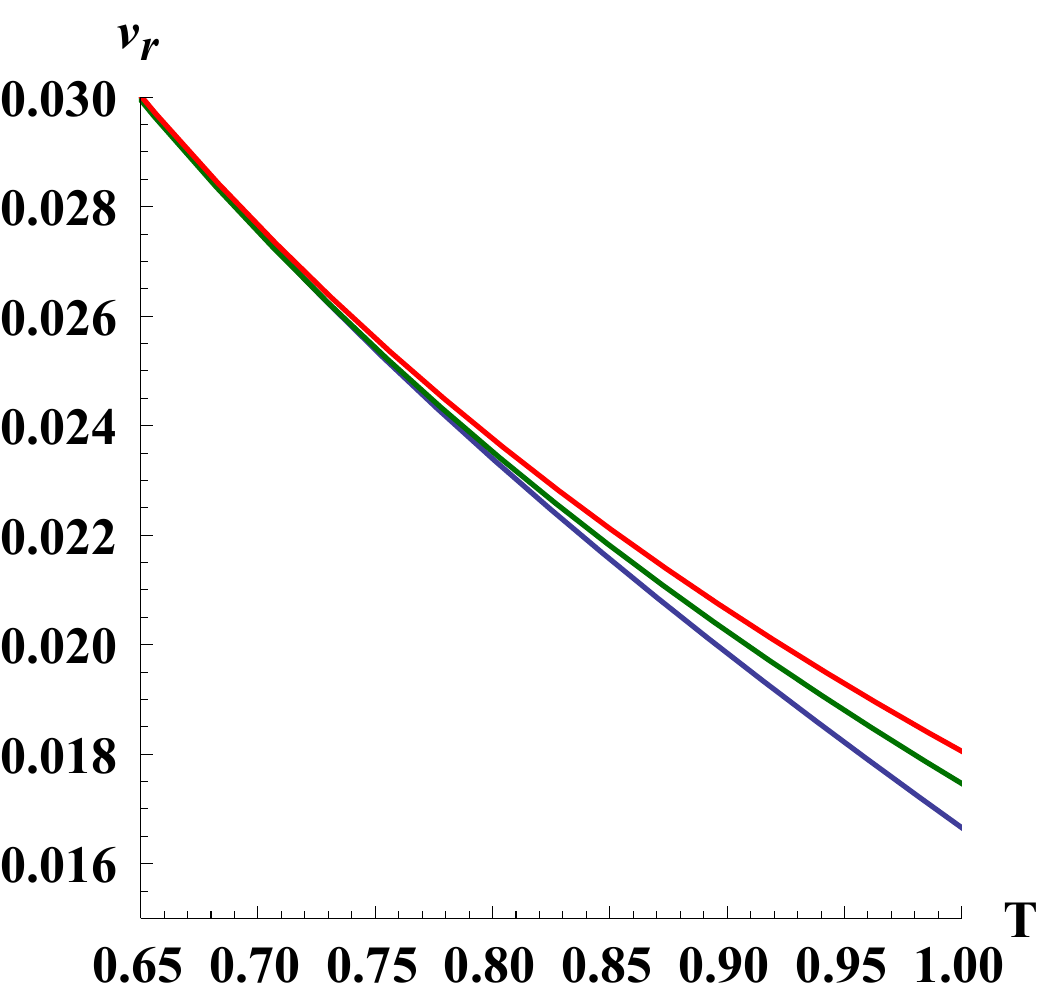}
\caption{The observable radial velocity $v_r$.  The constants are chosen as follows: $\beta = -0.111113$, $v_{r0}=0.05$, $\chi_{in}=0.084$. The indexes are $k = -1.01$ for the red line, $k = -1.5$ for the green line, and $k = -2.5$ for the blue line.}
\label{fig1}	
\end{minipage}	
	\hfill
\begin{minipage}{0.48\linewidth}\centering
 		\includegraphics[width=0.9\linewidth]{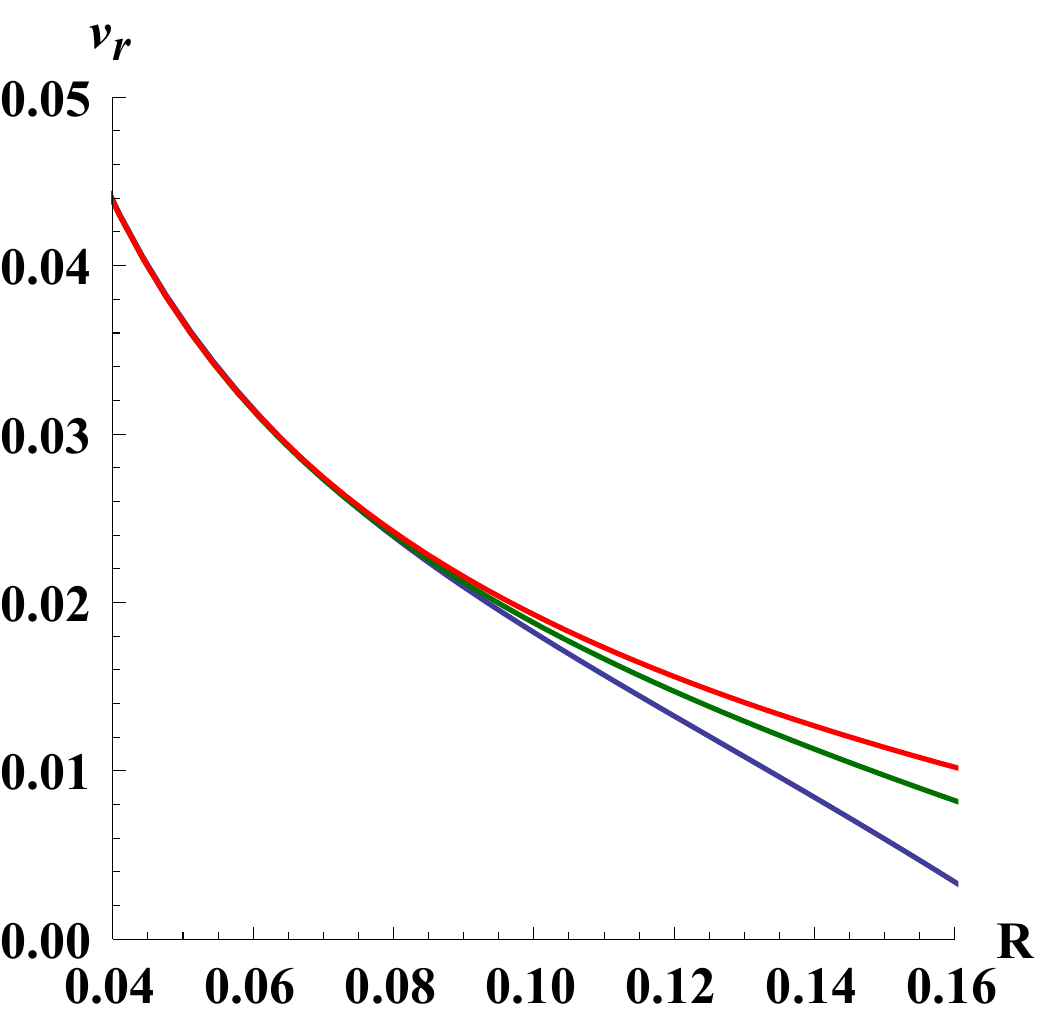}
	\caption{The dependence $v_r(R)$ under $\beta = -0.111113$, $v_{r0}=0.05$, $\chi_{in}=0.084$. $k = -1.01$ for the red line, $k = -1.5$ for the green line, and $k = -2.5$ for the blue line.}
\label{fig2}
\end{minipage}	
\end{figure}

First, we consider a purely radial motion ($u^3=0$). Integrating (\ref{geq_u0}) together with \mbox{(\ref{geq_u1})} numerically, we find the observable radial velocity $v_r$ of the test particle presented in Figures \ref{fig1} and \ref{fig2}. Figure \ref{fig1} shows how $v_r$ changes with time for different values of the exponent $k$. The time variable is parametrized by the function $T(t)$. The current moment of time corresponds to $T=1$. We choose the initial conditions in the following way: $T_{in}$ indicates the moment when observations start, $\chi_{in}$ and $v_{r0}$ are the initial position and the initial radial velocity of the test particle, respectively.

Figure \ref{fig2} demonstrates the dependence $v_r(R)$, where $R$ is the dimensionless observable radial distance to the particle from the center of symmetry:
\begin{equation}\label{R}
R\equiv\frac{1}{a_0}\int_0^{\chi}r(t,\chi)\mathrm{d}\chi=\frac{2 T^{1+\frac{k}{2}}}{\sqrt{|\beta|}} \arctanh (T^{-\frac{k}{2}}\frac{\chi}{2}\sqrt{|\beta|}).
\end{equation}
In view of (\ref{a0}) and (\ref{r0}), the current size of the universe in such dimensionless units is $R_0=r_0/a_0=2.78$.

\begin{figure}[t]
\centering
	\begin{minipage}{0.48\linewidth}\centering
 		\includegraphics[width=0.9\linewidth]{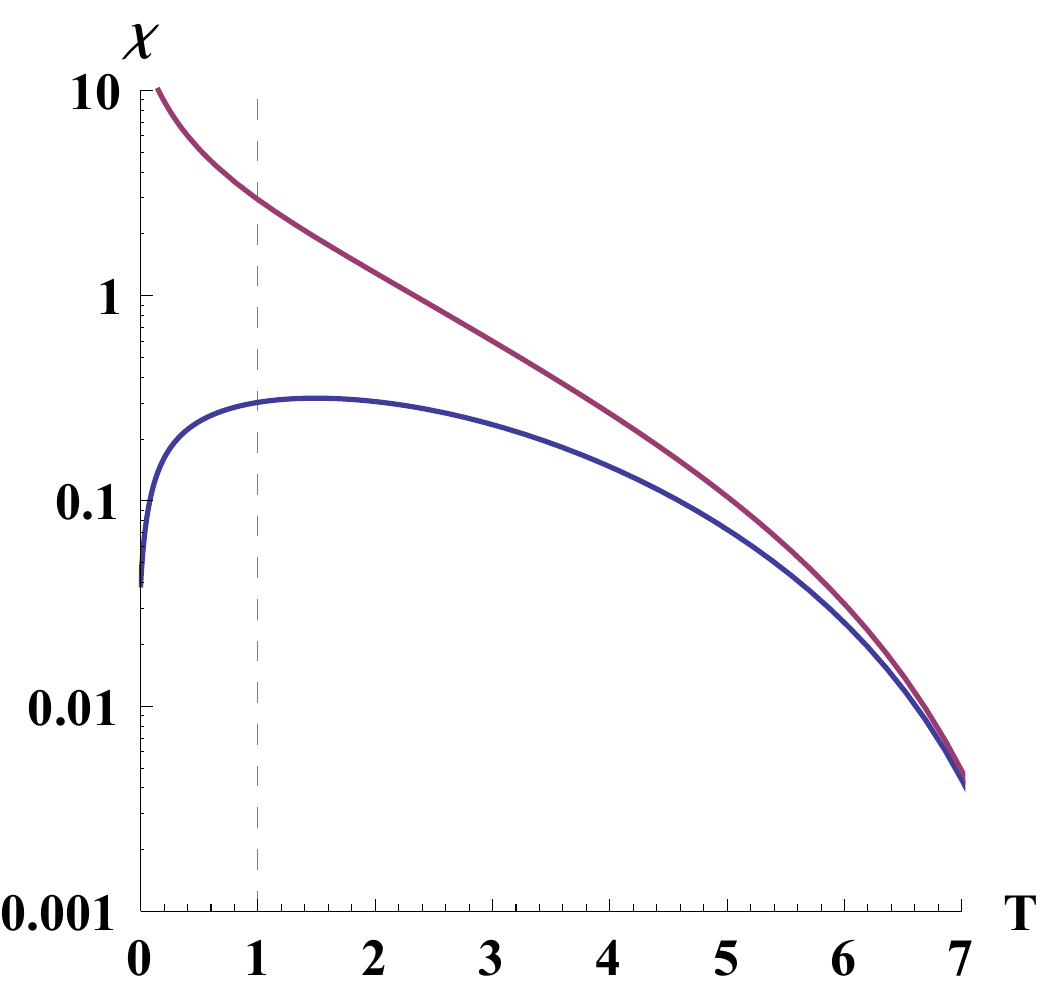}
\caption{Relative positions of the singularity (\ref{sing}) (the violet line) and $\chi(T)$ (the blue line). The dashed line indicates the current age of the universe. The constants are  $\beta = -0.111113$, $v_{r0}=0.05$, $\chi_{in}=0.084$, $k = -2.5$.}
\label{fig3}	
	\end{minipage}	
\hfill
	\begin{minipage}{0.48\linewidth}\centering
 		\includegraphics[width=0.9\linewidth]{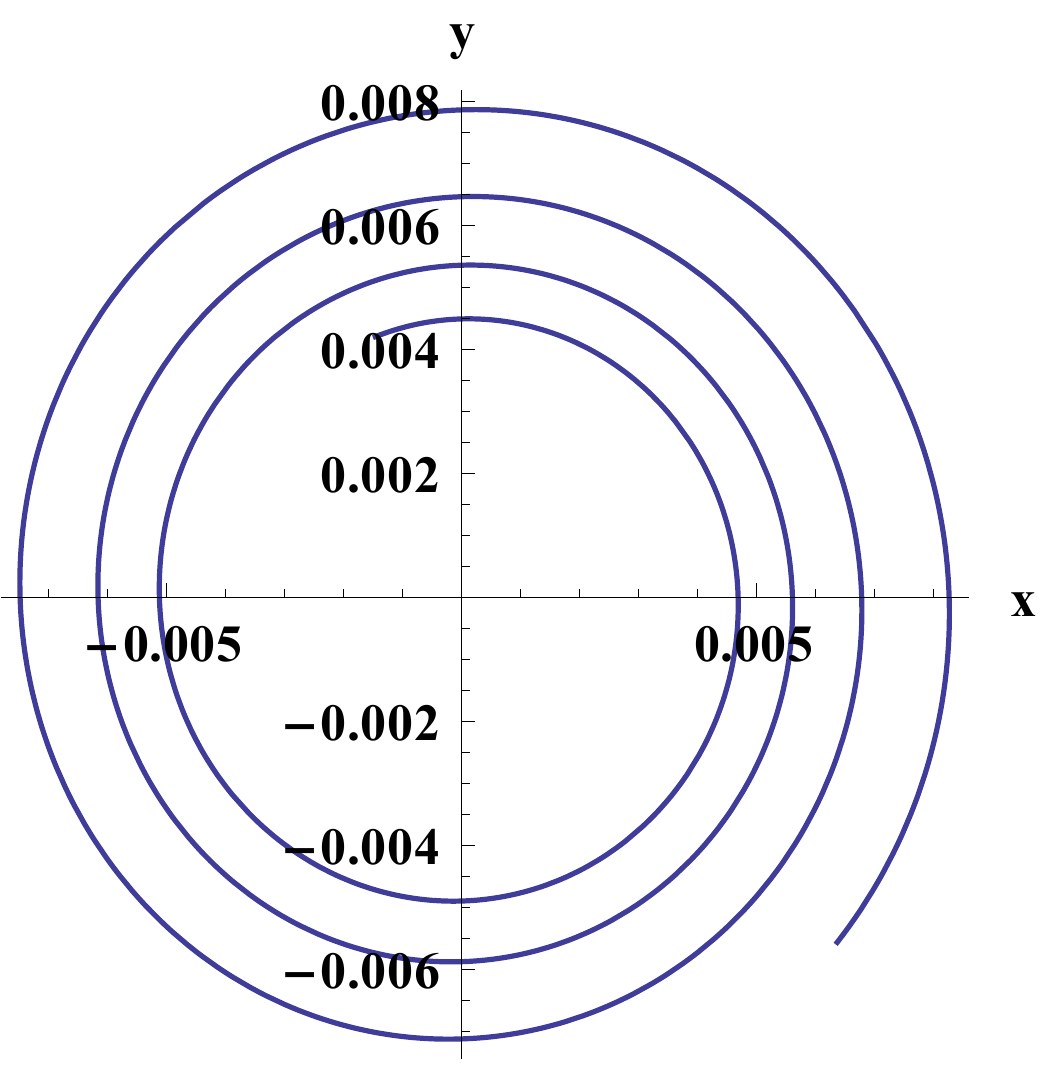}
	\caption{The spiralling out trajectory of the test particle in the case $\chi = \mathrm{const}$. $x$ and $y$ are ordinary Cartesian coordinates measured in the units of $a_0$ (\ref{a0}). The time interval between the start and finish points is taken here to be $\Delta T=0.45$ that corresponds to $\Delta \tau\approx 7 \times 10^{16} \mathrm{sek}$.}
	\label{fig4}
	\end{minipage}
\end{figure}

Figure \ref{fig3} shows the relative positions of the singularity (\ref{sing}) and the solution $\chi(T)$ of the geodesic equations (\ref{geq_u0}), (\ref{geq_u1}).

It is seen that the radial velocity decreases slightly with time and radial distance, and there are no specific features in its behavior up to the current moment of time. From Figure \ref{fig3} it follows that within the whole range of the solution the singularity remains indistinguishable for test particle. 

In case of $\chi=\mathrm{const}$ the geodesic equations can be integrated explicitly. The Equations~(\ref{geq_u0}) and (\ref{geq_u3}), respectively, give
\begin{equation}\label{u0rot}
\frac{\mathrm{d}u^0}{\mathrm{d}s}=-\frac{1}{2g_{00}}\left((u^0)^2 \dot{g}_{00}+2(u^3)^2\chi^2r\dot{r}\right),
\end{equation}
\begin{equation}\label{u3rot}
\frac{\mathrm{d}u^3}{\mathrm{d}s}=-\frac{2u^0u^3\dot{r}}{r}.
\end{equation}

The Equation (\ref{u3rot}) rewritten in the form
\begin{equation}
\frac{1}{u^3}\frac{\mathrm{d}u^3}{\mathrm{d}t}=-\frac{2\dot{r}}{r},
\end{equation}
which immediately leads to the angular momentum conservation law
\begin{equation}\label{u33}
u^3=\frac{L}{r^2\chi^2},
\end{equation}
where $L$ is a constant of integration. The normalization condition (\ref{geq_int}) gives
\begin{equation}\label{introt}
1-\frac{r^2\chi^2}{g_{00}}\frac{(u^3)^2}{(u^0)^2}=\frac{1}{g_{00}(u^0)^2},
\end{equation}
that, combined with (\ref{u33}), results in the expression
\begin{equation}\label{u00}
u^0=\pm\frac{\sqrt{L^2+r^2\chi^2}}{\sqrt{g_{00}}r\chi}.
\end{equation}

According to the definition, regarding (\ref{u33}) and (\ref{u00}), the observable orbital velocity of the test particle reads
\begin{equation}\label{vfi}
v_{\varphi}=\frac{r\chi u^3}{\sqrt{g_{00}}u^0}=\pm\frac{L}{\sqrt{L^2+r^2\chi^2}}.
\end{equation}

Choosing the initial conditions as in the previous manner, denoting with $v_{\varphi0}$ the initial orbital velocity of the test particle, we find the constant $L$ regarding (\ref{vfi})
\begin{equation}
L=\frac{v_{\varphi0}\chi_{in}r_{in}}{\sqrt{1-v^2_{\varphi0}}}.
\end{equation}

Figure \ref{fig4} shows the trajectory of the test particle in the case $\chi=\mathrm{const}$. It can be found in standard way after numerical integration of the Equation (\ref{vfi}). In this graph, one should multiply the dimensionless distance by $a_0=1.58\times 10^{26}~\mathrm{m}$ to estimate its value in meters. Although $\chi$ remains constant, the ``polar'' radius of the trajectory $r(t,\chi)\chi$ changes with time, and the test particle spirals around the center. 

\begin{figure}[t]
\centering
	\begin{minipage}{0.48\linewidth}\centering
 		\includegraphics[width=0.9\linewidth]{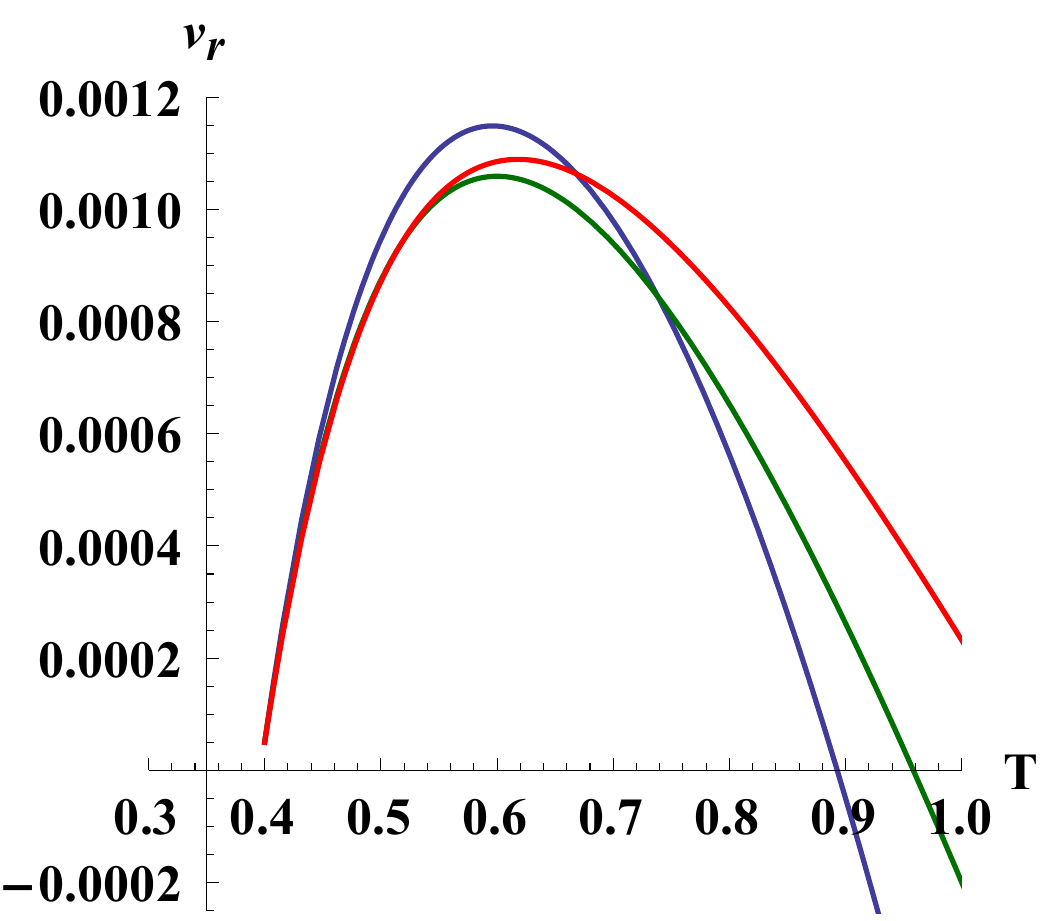}
\caption{The observable radial velocity in the general case of motion.  The constants are chosen as follows: $\beta = -0.111113$, $v_{r0}=0.00005$, $L = 0.001$, $\chi_{in}=0.084$, $k = -1.01$ (the red line), $k = -1.5$ (the green line) and $k = -2.5$ (the blue line).}
\label{fig5}	
	\end{minipage}	
\hfill
	\begin{minipage}{0.48\linewidth}\centering
 		\includegraphics[width=0.9\linewidth]{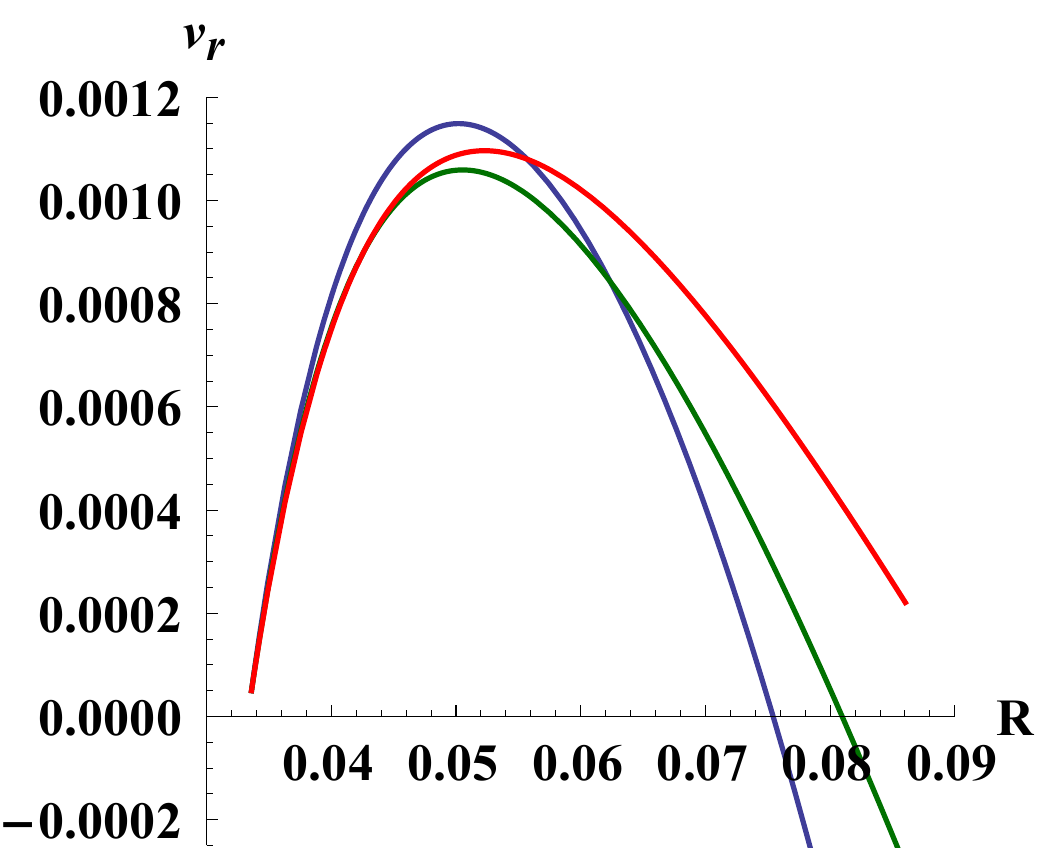}
\caption{The dependence $v_r(R)$ in the general case of motion. The constants are chosen as follows: $\beta = -0.111113$, $v_{r0}=0.00005$, $L = 0.001$, $\chi_{in}=0.084$, $k = -1.01$ (the red line), $k = -1.5$ (the green line) and $k = -2.5$ (the blue line).}
\label{fig6}	
	\end{minipage}
\end{figure}

\begin{figure}[t]
\centering
	\begin{minipage}{0.48\linewidth}\centering
 \includegraphics[width=1\linewidth]{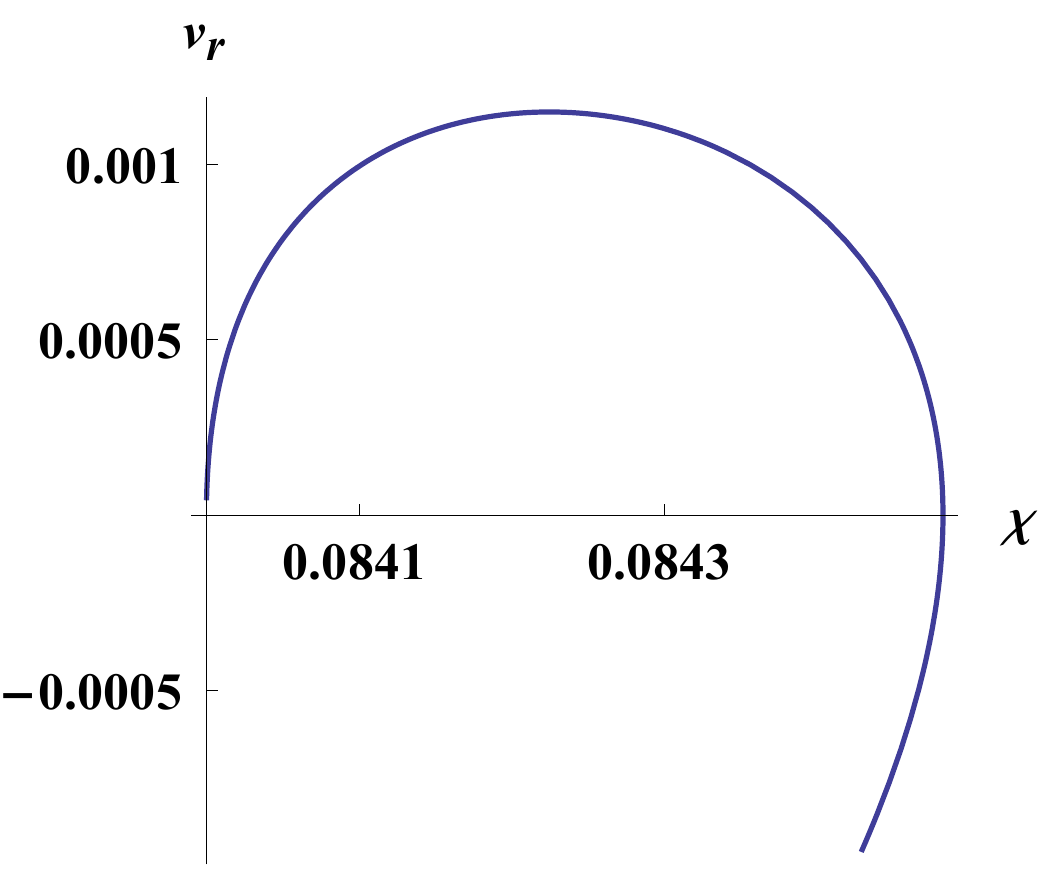}
\caption{The dependence $v_r(\chi)$ in the general case of motion. The constants are $\beta = -0.111113$, $v_{r0}=0.00005$, $L = 0.001$, $\chi_{in}=0.084$, $k = -2.5$.}
\label{fig7}	
	\end{minipage}	
\hfill
	\begin{minipage}{0.48\linewidth}\centering
 		\includegraphics[width=0.9\linewidth]{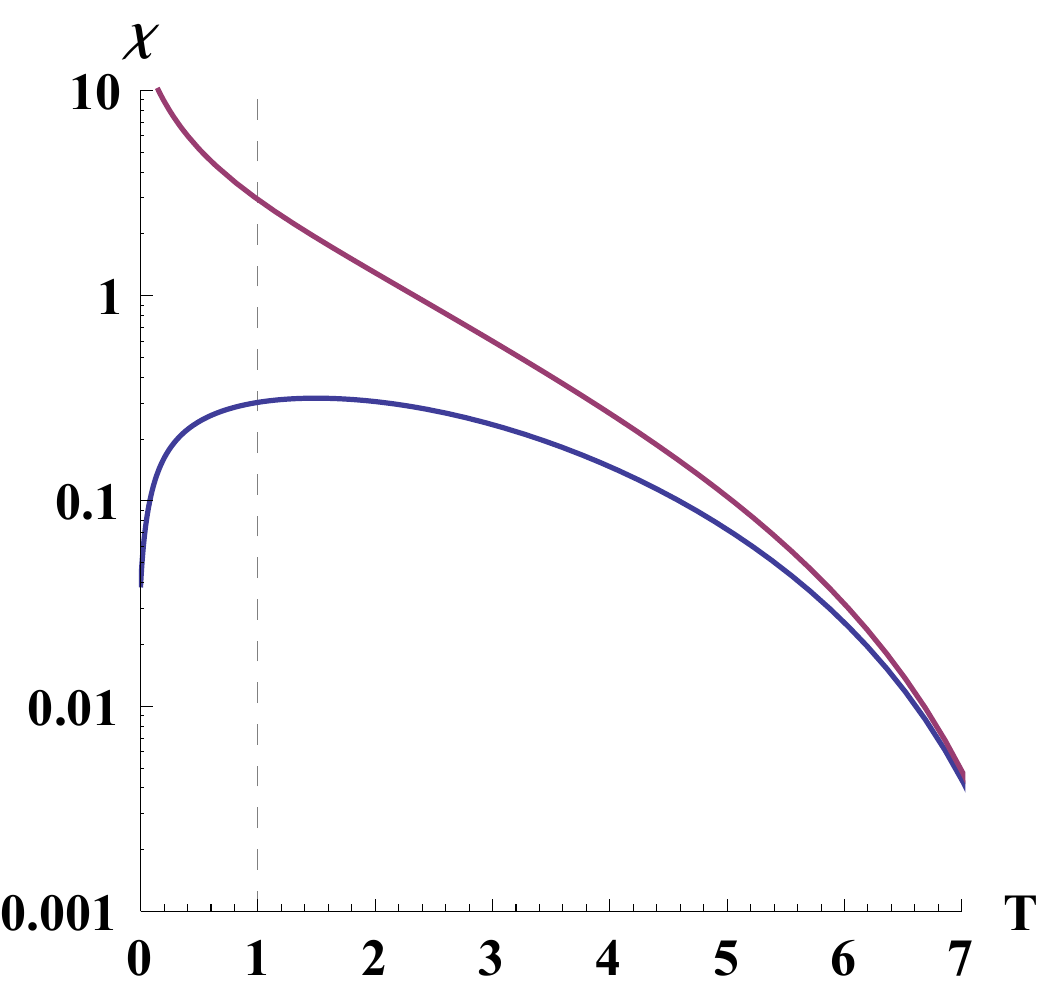}
\caption{Relative positions of the singularity (\ref{sing}) (the violet line) and $\chi(T)$ (the blue line) in the general case. The dashed line indicates the current age of the universe. The constants are $\beta = -0.111113$, $v_{r0}=0.00005$, $L = 0.001$, $\chi_{in}=0.084$, $k = -2.5$. }
\label{fig8}	
	\end{minipage}
\end{figure}

In the general case, the motion of the test particle can be analyzed with respect to Figures \ref{fig5}--\ref{fig8}. 

Figure \ref{fig5} shows the observable radial velocity of the test particle when both $u^1$ and $u^3$ are nonzero. $L$ is the dimensionless angular momentum defined by
\begin{equation}
L=\frac{T_{in}\chi_{in}}{1-T^{-k}_{in}\left(\frac{\chi_{in}}{2}\right)^2 |\beta|}\sqrt{1-v^2_0},
\end{equation}
where $v_0=\sqrt{v^2_{r0}+v^2_{\varphi0}}$ is the initial observable particle velocity.

The dependence $v_r(R)$ is presented in Figure \ref{fig6} built in accordance with the numeric solution of the geodesic equations. It may be treated as a phase portrait, from where we see that there exists a turning point. Moreover, at some moments, the radial velocity changes its sign. The interesting thing in this model is that the observable distance continues increasing even under negative $v_r$. This is caused by the fact that $R$ depends both on $\chi$ and $t$, it can still grow even when $\chi$ decreases. This is reflected in Figure \ref{fig7}, where $\chi$ decreases under negative $v_r$.

Figure \ref{fig8} again demonstrates that the singularity (\ref{sing}) is indistinguishable.

\section{Null Geodesics}\label{Sec4}
The geodesic equations for the massless particles have the same form as (\ref{geq_u0})--(\ref{geq_u3}), but the derivative of the components of the four-velocity must be taken with respect to the affine parameter $\lambda$ and the normalization equation $u^{\mu}u_{\mu}=0$ gives
\begin{equation}\label{u000}
(u^0)^2= \frac{r^2}{g_{00}}(u^1)^2+\frac{r^2\chi^2}{g_{00}}(u^3)^2.
\end{equation}

The radial motion ($u^3=0$) of the light rays is of particular interest because the solution of geodesic equations in this case is used to calculate the luminosity distance and hence the redshift--magnitude relation. Putting $u^0$ from (\ref{u000}) into (\ref{geq_u1}), we obtain the differential equation with separable variables and integrate it with the following result\vspace{6pt}
\begin{equation}
u^1=4 a_0^k T^k \frac{\left(1- T^{-k}|\beta|\left(\frac{\chi }{2}\right)^2 \right)^2}{1-T^{-k}(1+k)|\beta| \left(\frac{\chi }{2}\right)^2 }\left(\frac{1+\frac{\chi}{2} \sqrt{|\beta|} T^{-\frac{k}{2}}}{1-\frac{\chi}{2} \sqrt{|\beta| } T^{-\frac{k}{2}}}\right)^{\frac{2 T^{k/2}}{\sqrt{|\beta|}} \sqrt{T^{-1} + T^{-k}|\beta|}}, 
\end{equation}
and hence due to (\ref{u000})
\begin{equation}
u^0=4a_0^{k+1} T^{k+1} \frac{\left(1- T^{-k}|\beta|\left(\frac{\chi }{2}\right)^2 \right)^2}{\left(1-T^{-k}(1+k)|\beta| \left(\frac{\chi }{2}\right)^2\right)^2 }\left(\frac{1+\frac{\chi}{2} \sqrt{|\beta|} T^{-\frac{k}{2}}}{1-\frac{\chi}{2} \sqrt{|\beta| } T^{-\frac{k}{2}}}\right)^{\frac{2 T^{k/2}}{\sqrt{|\beta|}} \sqrt{T^{-1} + T^{-k}|\beta|}}.
\end{equation}

The detailed procedure of obtaining the redshift--magnitude relation in this model can be found in \cite{Kopteva19}.

Equations (\ref{geq_u0})--(\ref{geq_u3}) and (\ref{u000}) can also be explicitly integrated under $\chi=\mathrm{const}$. Thus, \mbox{(\ref{geq_u3})} immediately gives 
\begin{equation}
u^3=\frac{b}{\chi^2 r^2},
\end{equation}
where $b$ is a constant of integration, with the sense of the impact parameter.
Using (\ref{u000}) we~find 
\begin{equation}
\frac{\mathrm{d}\varphi}{\mathrm{d}t}=\frac{1-T^{-k}\left(\frac{\chi}{2}\right)^2(1+k)|\beta|}{\chi a_0 T}.
\end{equation}

Changing the variable in the previous way, we obtain
\begin{equation}
\varphi=\int\frac{1-T^{-k}\left(\frac{\chi}{2}\right)^2(1+k)|\beta|}{\chi T} \frac{1}{\sqrt{T^{-1}+|\beta|T^{-k}}}\mathrm{d}T,
\end{equation}
that can be explicitly integrated in terms of hypergeometric function $\mathrm{{}_{2}F_{1}}$.

\begin{align}\label{ppp}
\varphi+\varphi_0&=\frac{1}{2k\chi\sqrt{T}}4T\,\mathrm{{}_{2}F_{1}}\left(\frac{1}{2}, \frac{1}{2-2k}; 1 + \frac{1}{2-2k}; -T^{1-k}|\beta|\right)+\\ \nonumber
&+(1+k)\sqrt{1+T^{1-k}|\beta|}\chi^2\left(1-\frac{T^{k-1}}{(2-k)|\beta|} \,\mathrm{{}_{2}F_{1}}\left(1, 1+\frac{1}{2-2k}; \frac{4-3k}{2-2k}; -\frac{T^{k-1}}{|\beta|}\right)\right)
\end{align}

\begin{figure}[t]
\centering
	\begin{minipage}{0.48\linewidth}\centering
 \includegraphics[width=0.9\linewidth]{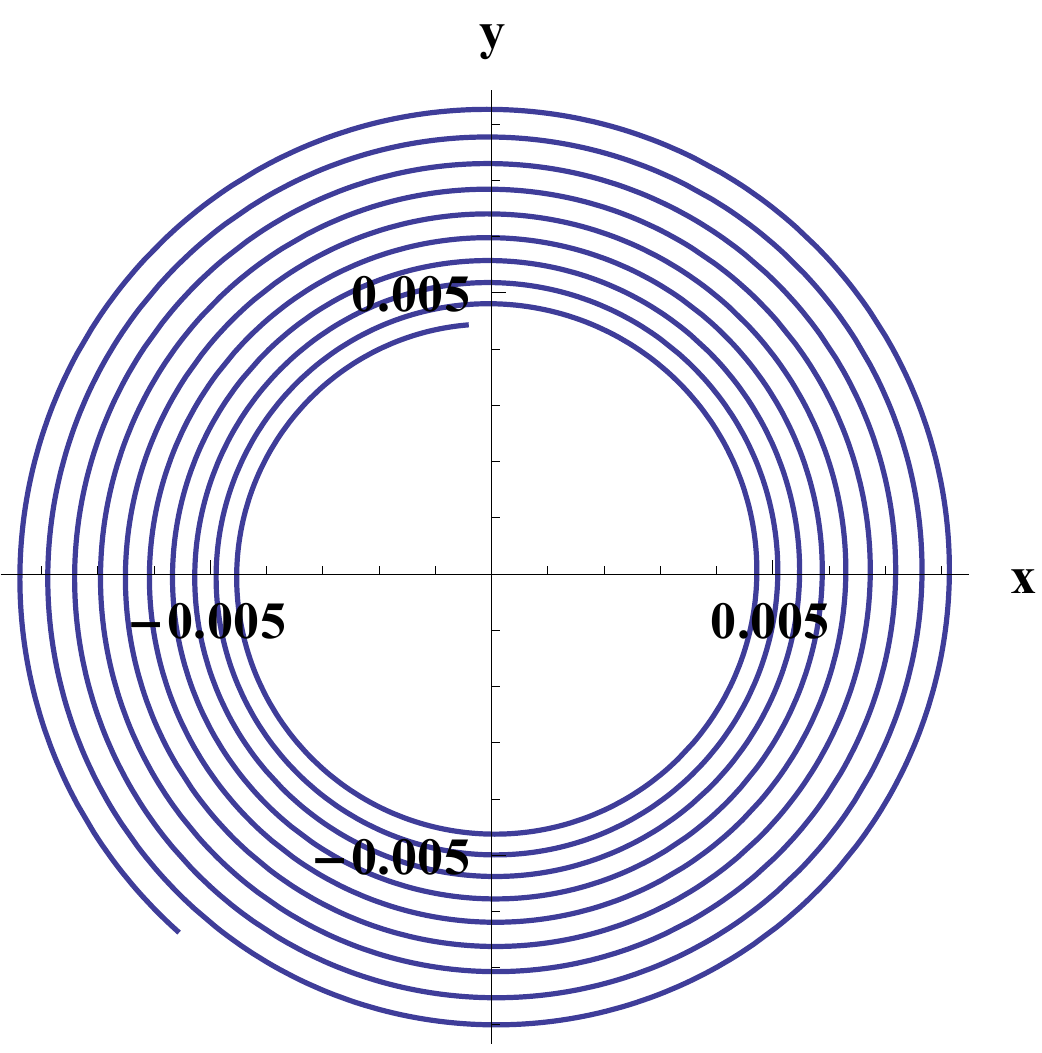}
	\caption{The trajectory of photons under $\chi=\mathrm{const}=0.009$. $\Delta T =0.45$, $k=-2.05$. $x$ and $y$ are dimensionless Cartesian coordinates in the plane $\theta=\pi/2$. To restore the distance, one should multiply its dimensionless value by $a_0$.}
\label{fig9}
	\end{minipage}	
\hfill
	\begin{minipage}{0.48\linewidth}\centering
 		\includegraphics[width=0.9\linewidth]{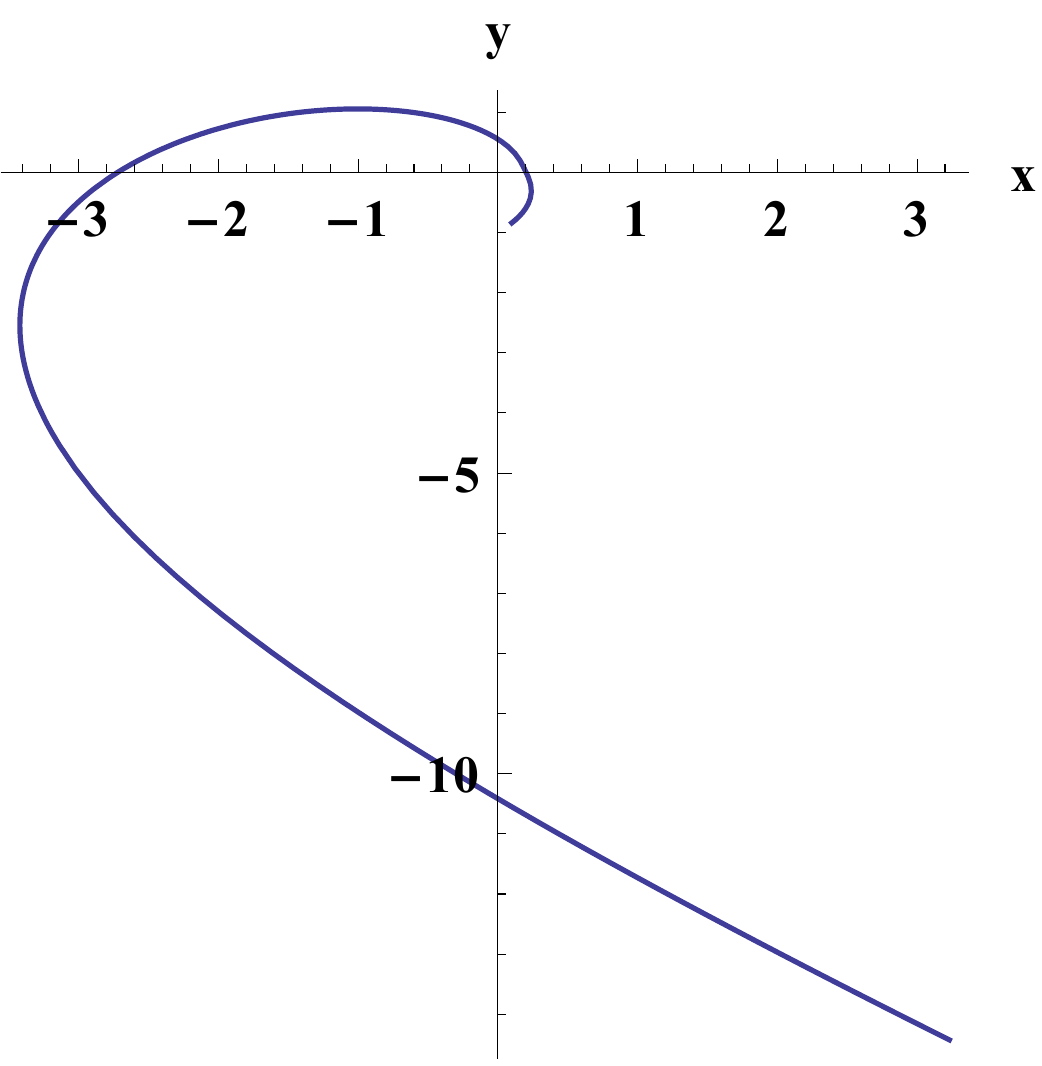}
	\caption{The trajectory of photons in the general case. The constants are chosen as follows $k=-2.1$, $b=0.9$, $\varphi_{in}=\chi_{in}=0.22$, $u^3_{in}=0.08$.}
\label{fig10}	
	\end{minipage}
\end{figure}

Figure \ref{fig9} shows the trajectory of photons according to (\ref{ppp}). The radius of the two-sphere $\chi=\mathrm{const}$ changes with time, so that the photons are spiralling out of the center. 

In the general case, numerical integration of (\ref{geq_u1}) together with (\ref{geq_u3}) gives the photon trajectory shown at Figure \ref{fig10}.

\begin{figure}[H]\centering
 		\includegraphics[width=0.48\linewidth]{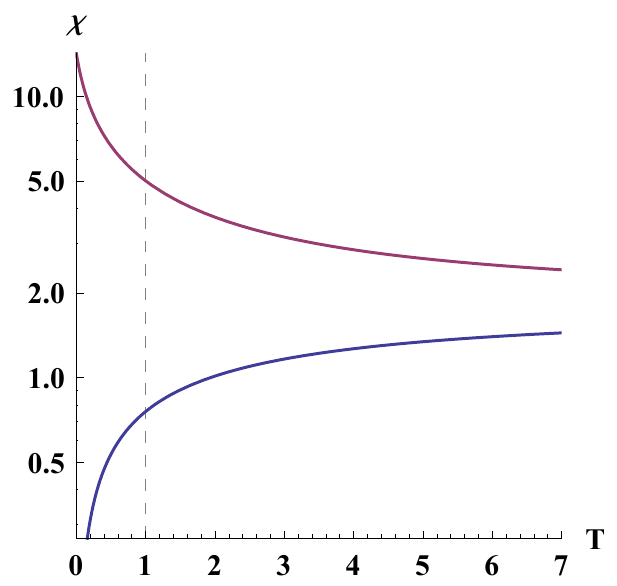}
	\caption{Relative positions of the singularity (\ref{sing}) (the violet line) and $\chi(T)$ (the blue line) in the general case of photon motion. The dashed line indicates the current age of the universe. The constants are $\beta = -0.111113$, $L = 0.0019$, $\chi_{in}=0.039$, $k = -2.5$.}
\label{fig11}
\end{figure}

From Figure \ref{fig11} it follows that, up to the current moment of time, the singularity is also unreachable for null geodesics. 

\section{Conclusions}\label{Con}
Despite the singularities of the Stephani cosmological models, some of them were found to demonstrate appropriate features for application to the real universe. For such a model, we investigate the main scenarios of the test particle and photon motion.  In the case of purely radial motion, the radial velocity decreases slightly with time and radial distance. There are no specific features in its behavior up to the current moment in time. The motion under constant radial coordinate appears to spiral out of the center both for particles and photons. The interesting thing in this model is that, in the case of the test particle motion with arbitrary initial velocity, the observable distance $R$ increases even under negative $v_r$, which is caused by the fact that $R$ depends on both $T$ and $\chi$ (see (\ref{R})), so it can grow even when $\chi$ decreases. The singularities are indistinguishable for observations and do not influence the test particles and photons motion up to the current age of the universe as well as in the far enough future. Analysis of the geodesic structure with respect to the singularity behavior shows that the closer the exponent $k$ to $-1$, the slower the solution $\chi(T)$ tends towards singularity. Figures \ref{fig3}, \ref{fig8} and \ref{fig11} are built for the concrete value of $k=-2.5$; however, it does not restrict the generality of conclusions. Furthermore, it is possible to match the Stephani universe with another suitable cosmological model (e.g., the de Sitter one, as in \cite{Korkina13}) to exclude the singularities at all.

\section*{Acknowledgements}
The authors acknowledge the Research Centre for Theoretical Physics and Astrophysics, Institute of Physics, Silesian University in Opava for institutional support. This research was funded by the Research Centre for Theoretical Physics and Astrophysics.

\end{document}